\documentstyle[12pt,epsf]{article}
\def\PR #1 #2 #3 {Phys.~Rev.~{\bf #1}, #2 (#3)}
\def\PRL #1 #2 #3 {Phys.~Rev.~Lett.~{\bf #1}, #2 (#3)}
\def\PRD #1 #2 #3 {Phys.~Rev.~D~{\bf #1}, #2 (#3)}
\def\PLB #1 #2 #3 {Phys.~Lett.~B~{\bf #1}, #2 (#3)}
\def\NPB #1 #2 #3 {Nucl.~Phys.~{\bf B#1}, #2 (#3)}
\def\RMP #1 #2 #3 {Rev.~Mod.~Phys.~{\bf #1}, #2 (#3)}
\def\ZPC #1 #2 #3 {Z.~Phys.~C~{\bf #1}, #2 (#3)}
\def\CPC #1 #2 #3 {Comp.~Phys.~Comm.~{\bf #1}, #2 (#3)}

\begin{document}

\rightline{hep-ph/9807340}
\medskip
\rightline{July 1998}
\bigskip\bigskip

\begin{center} {\Large \bf Single-top-quark production at hadron colliders}
\\
\bigskip\bigskip\bigskip\bigskip
{\large\bf T.\ Stelzer, Z.\ Sullivan} {\large and}
{\large\bf S.\ Willenbrock} \\ \medskip Department of Physics \\
University of Illinois \\ 1110 West Green Street \\  Urbana, IL\ \ 61801 \\
\bigskip
\end{center}
\bigskip\bigskip\bigskip

\begin{abstract}
Single-top-quark production probes the charged-current weak
interaction of the top quark, and provides a direct measurement of the
CKM matrix element $V_{tb}$.  We perform two independent analyses to
quantify the accuracy with which the $W$-gluon fusion ($gq\to t\bar
bq$) and $q\bar q \to t\bar b$ signals can be extracted from the
backgrounds at both the Tevatron and the LHC.  Although perturbation
theory breaks down at low transverse momentum for the $W$-gluon fusion
$\bar b$ differential cross section, we show how to obtain a reliable
cross section integrated over low $\bar b$ transverse momenta up to a
cutoff.  We estimate the accuracy with which $V_{tb}$ can be measured
in both analyses, including theoretical and statistical uncertainties.
We also show that the polarization of the top quark in $W$-gluon
fusion can be detected at the Fermilab Tevatron and the CERN LHC.
\end{abstract}

\addtolength{\baselineskip}{9pt}

\newpage

\section{Introduction}

\indent\indent Single-top-quark production at the Fermilab Tevatron
and the CERN Large Hadron Collider (LHC) provides an opportunity to
study the charged-current weak-interaction of the top quark
\cite{DW,Y,EP,CY,HBB,CP,SW}.  Within the standard model,
single-top-quark production offers a means to directly measure the
Cabibbo-Kobayashi-Maskawa matrix element $V_{tb}$.  Beyond the
standard model, it is sensitive to a non-standard $Wtb$ vertex, and to
exotic single-top-quark production processes involving new particles
\cite{CY,CMY,ABES,DZ,LOY,S,LCHZX,DYYZ,TY}.  In order to be a useful
probe, the measurement of single-top-quark production must be
accompanied by an accurate calculation of the standard-model
production cross section and experimental acceptance, as well as an
analysis of the associated backgrounds.

\begin{figure}
\begin{center}
\leavevmode
\epsfxsize= 3.0in
\epsfbox{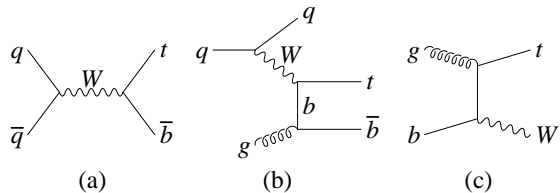} 
\end{center}
\caption{Representative Feynman diagrams for single-top-quark
production at hadron colliders: (a) $s$-channel production,
(b) $t$-channel production ($W$-gluon fusion), and (c) associated 
production with a $W$ boson.\label{Feynman}}
\end{figure}

It is useful to distinguish between three different types of
single-top-quark production, based on the virtuality of the
$W$~boson. Fig.~\ref{Feynman}(a) shows the leading-order Feynman
diagram for $s$-channel single-top-quark production \cite{CP,SW}. This
process has the theoretical advantage of proceeding via
quark-antiquark annihilation, so the partonic flux can be constrained
from Drell-Yan data \cite{DPZ}.  The next-to-leading-order calculation
has been performed for this channel \cite{SmithW}, as well as a study
of the acceptance and backgrounds \cite{SW,tev2000}.
Fig.~\ref{Feynman}(b) shows a Feynman diagram for $t$-channel
single-top-quark production, often referred to as $W$-gluon fusion
\cite{DW,Y,EP,CY,HBB}.\footnote{The $s$-channel process is sometimes
referred to as the $W^*$ process; however, the $W$ boson in the
$t$-channel process is also off-shell.}  The primary advantage of this
channel is statistics.  The cross section is almost 3 times larger
than that of the $s$-channel process at the Tevatron ($\sqrt S = 2$
TeV), and the cross section at the LHC is 100 times larger than at the
Tevatron.  The production cross section was recently calculated by us
at next-to-leading order \cite{BV,SSW}, and the acceptance and
backgrounds have been most completely studied in Ref.~\cite{tev2000}.
Fig.~\ref{Feynman}(c) shows a Feynman diagram for $Wt$ production,
where an on-shell $W$ is produced \cite{HBB,M}. This process proceeds
via a gluon-$b$ interaction, which makes the cross section negligible
at the Tevatron. However, at the LHC it contributes about 20\% of the
total single-top-quark cross section.  Neither the
next-to-leading-order cross section,\footnote{The
next-to-leading-order cross section is available for the identical
process of $Wc$ production \cite{GKL}.}  nor the calculation of the
acceptance and backgrounds for this process, are yet available.

In this paper we calculate the acceptance and backgrounds for
single-top-quark production via $W$-gluon fusion at the Tevatron and
LHC.  There are a number of differences with the analysis of
Ref.~\cite{tev2000}.  The most significant improvement is that we
perform an accurate calculation of the acceptance, using our
next-to-leading-order calculation of the total cross section.  This is
an essential ingredient in the extraction of the cross section from
experiment, and can be used to normalize any future studies.  The
acceptance cannot simply be calculated by comparing the cross section
from Fig.~\ref{Feynman}(b) with and without cuts, due to the breakdown
of perturbation theory in the region where the initial gluon splits
into a nearly-collinear $b\bar b$ pair.  The correct way to treat the
collinear region and calculate the acceptance is discussed in detail
in Section 2.

Our analysis of backgrounds differs from that of Ref.~\cite{tev2000}
in that we advocate the use of one and only one $b$ tag to isolate the
signal, while Ref.~\cite{tev2000} requires one or more $b$ tags.  The
main motivation for this is that we desire to separate
single-top-quark production via $W$-gluon fusion (which usually has
only the $b$ quark from top decay in the fiducial region) from the
$s$-channel process (which usually has a $b$ and a $\bar b$ in the
fiducial region).  This provides two independent measurements of
$V_{tb}$, with different backgrounds and theoretical uncertainties.
Perhaps more importantly, the two processes are generally influenced
by new physics in different ways, so the deviation of each process
from the standard model would be a useful diagnostic
\cite{CY,CMY,ABES,DZ,LOY,S,LCHZX,DYYZ,TY}.  We also perform an
analysis of $W$-gluon fusion at the LHC, while the study of
Ref.~\cite{tev2000} concentrated on the Tevatron.

Based on these results, we study the sensitivity with which $V_{tb}$
can be extracted from single-top-quark production via $W$-gluon fusion
at both the Tevatron and the LHC, taking into account both statistical
and theoretical uncertainties.  We also perform an analysis of the
$s$-channel process, and compare the results with those of the
$W$-gluon-fusion process.  We consider the data collected during Run I
at the Tevatron ($\sqrt S = 1.8$ TeV) from 1992--1995 (110 pb$^{-1}$),
the data that will be collected in Run II ($\sqrt S = 2$ TeV)
beginning in 2000 (2 fb$^{-1}$), and additional data which may be
collected (at the same energy) beyond Run II (30 fb$^{-1}$).

Since the top quark is produced via the weak interaction in
single-top-quark processes, it has significant polarization
\cite{CY,HBB}.  An optimal basis for the measurement of this
polarization, both for the $s$-channel process and for $W$-gluon
fusion, was recently introduced in Ref.~\cite{MP}.  We quantify the
integrated luminosity required to observe this polarization, including
the effects of acceptance and jet resolution.

The paper is organized as follows. In Section~2 we calculate the
acceptance for single-top-quark production via $W$-gluon fusion. We
pay particular attention to the issues associated with the splitting
of the initial gluon into a nearly-collinear $b\bar b$ pair.  In
Section~3 we briefly discuss our calculational techniques.  In
Section~4 we present results for the signal and backgrounds at the
Tevatron and the LHC, and analyze the accuracy with which $V_{tb}$ can
be extracted. Section~5 contains an analysis of the $s$-channel
process.  Section~6 is concerned with the polarization of the top
quark in single-top-quark processes.  We summarize our results in
Section~7.

\section{Acceptance}

\indent\indent We recently calculated the next-to-leading-order total
cross section for single-top-quark production via $W$-gluon fusion in
Ref.~\cite{SSW}.  The results are listed in Table~\ref{sigma}.
Experimentally, only the cross section which lies within the
geometrical acceptance of the detector is measurable, so it is
important to calculate this acceptance.  Normally this is
straightforward; one simply computes the ratio of the tree-level cross
section with and without cuts.  However, the total cross section for
$W$-gluon fusion cannot simply be calculated from
Fig.~\ref{Feynman}(b), because perturbation theory breaks down in the
region where the initial gluon splits into a nearly-collinear $b\bar
b$ pair.  Thus we must consider the correct way to calculate the
acceptance.

\begin{figure}
\begin{center}
\input{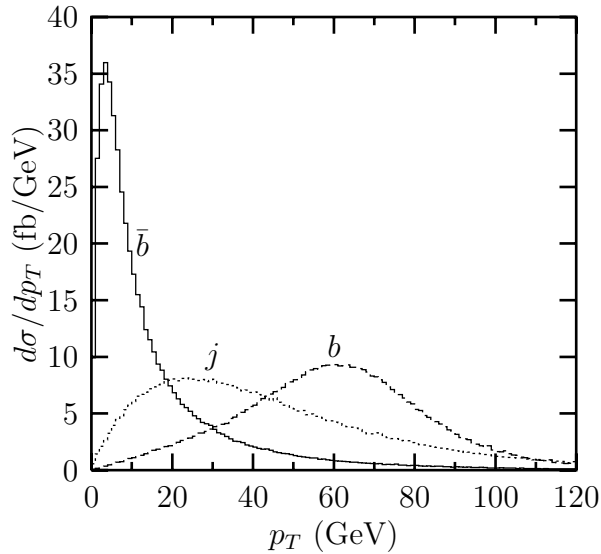}    
\end{center}
\caption{Transverse momentum distributions of the $\bar b$ antiquark
(solid line), the $b$ quark from top decay (dashed line), and the light-quark
jet (dotted line), in single-top-quark production via $W$-gluon fusion
($gq \to t\bar b q$) at the Tevatron ($\sqrt S=2$ TeV).\label{b_pt}}
\end{figure}

The $p_T$ spectrum of the $\bar b$ antiquark is shown in
Fig.~\ref{b_pt}.  It is peaked at small $p_T$, because the internal
$b$-quark propagator is close to being on shell when the initial gluon
splits into a nearly-collinear $b\bar b$ pair.  Since $d\sigma/dp_T^2
\sim 1/(p_T^2+m_b^2)$, the cross section with the $p_T$ of the $\bar
b$ antiquark above $p_{Tcut}$ is proportional to $\ln
[m_t^2/(p_{Tcut}^2+m_b^2)]$.  Another power of this logarithm appears
at every order in perturbation theory via collinear gluon radiation
from the internal $b$ quark, so the expansion parameter is $\alpha_s
\ln [m_t^2/(p_{Tcut}^2+m_b^2)]$.  Thus the calculation of the cross
section is more accurate the larger the choice of $p_{Tcut}$.

Unfortunately, it is not practical to simply choose a large value of
$p_{Tcut}$ and measure the cross section for $Wb\bar bj$ ($j$ denotes
the light-quark jet from the emission of the $t$-channel $W$ boson in
Fig.~\ref{Feynman}(b); $Wb$ are the decay products of the $t$ quark).
There is a large background from $t\bar t$ production, which yields
the final state $WWb\bar b$; this mimics the signal when the
additional $W$ boson decays to two jets, and one jet is missed.  To
suppress this background we search for the signal in the final state
$Wbj$, {\it i.e.}, we demand that the $\bar b$ antiquark {\em not}
appear in the final state.

Fortunately, the cross section with the $\bar b$ antiquark below
$p_{Tcut}$ can be calculated with good accuracy, provided $p_{Tcut}$
is sufficiently large.  This is achieved via a two-step procedure.  In
Ref.~\cite{SSW} we calculated the total cross section ($p_{Tcut} = 0$)
and summed the logarithmically-enhanced terms, $\alpha_s^n\ln^n
(m_t^2/m_b^2)/n!$, to all orders \cite{CY,HBB,OT,BHS,YC}.  To
calculate the cross section with the $\bar b$ antiquark {\em below}
$p_{Tcut}$, we simply take the total cross section and subtract from
it the cross section with the $\bar b$ antiquark {\em above}
$p_{Tcut}$:\footnote{The cross section with the $\bar b$ antiquark
above $p_{Tcut}$ is calculated using the scale $\mu^2 = p_T^2 + m_b^2$
in the gluon distribution function and the strong coupling.}
\begin{equation}
\sigma(p_T < p_{Tcut}) = \sigma_{NLO} - \sigma(p_T > p_{Tcut})\;.
\end{equation}
This is tantamount to integrating the transverse momentum of 
the $\bar b$ antiquark over all momenta below $p_{Tcut}$.  

We give in Table~\ref{sigma} the cross section for single-top-quark
production via $W$-gluon fusion with the $\bar b$ antiquark below
$p_{Tcut}=20$ GeV.  These numbers can be used to normalize future
studies.  For example, Ref.~\cite{tev2000} studied the signal in the
final state $Wbq$, using the process $qb \to qt$ to approximate the
$W$-gluon-fusion process, and normalizing to the total cross section.
However, it is more accurate to normalize to the cross section with
the $\bar b$ antiquark below some chosen $p_{Tcut}$ (20 GeV in
Ref.~\cite{tev2000}).\footnote{Ref.~\cite{tev2000} normalized to a
cross section of 1.6 pb; we see from Table~\ref{sigma} that a more
accurate cross section is 1.90 pb.}  HERWIG and PYTHIA\footnote{PYTHIA
uses backwards evolution of the initial-state $b$ distribution
function to give the initial $g \to b \bar b$ splitting.} also
simulate single-top-quark production via $W$-gluon fusion using $qb
\to qt$.

\begin{table}
\begin{center}
\caption{Cross sections (pb) for single-top-quark production via
$W$-gluon fusion with $m_t=$ 175 GeV.  The next-to-leading-order total
cross section is taken from Ref.~\protect\cite{SSW}. The last column
gives the cross section with the $\bar b$ antiquark below
$p_{Tcut}=20$ GeV.  The uncertainty is estimated from the scale
variation of the cross section, and does not include the uncertainty
in the parton distribution functions nor the uncertainty in the
top-quark mass.\label{sigma}}
\bigskip 
\begin{tabular}{ccc}
$\sqrt S$ & $\sigma_{NLO}$ & $\sigma(p_{T} < 20$ GeV) \\
\\
1.8 TeV $p\bar p$ & 1.70 $\pm$ 0.09 & 1.32 $\pm$ 0.14 \\
2 TeV $p\bar p$ & 2.44 $\pm$ 0.12 & 1.86 $\pm$ 0.20 \\
14 TeV $pp$ & 245 $\pm$ 12 & 164 $\pm$ 14 \\
\end{tabular} \end{center} 
\end{table}

Our strategy is therefore as follows.  We use the process in
Fig.~\ref{Feynman}(b) to calculate the differential cross section for
single-top-quark production via $W$-gluon fusion, using the scale
$\mu^2 = p_T^2+m_b^2$ in the gluon distribution function and the
strong coupling.\footnote{We use the scale $\mu^2 = Q^2$, where $Q^2$
is the virtuality of the $t$-channel $W$ boson, for the light-quark
distribution function \cite{SSW}.}  If the $p_T$ of the $\bar b$
antiquark is below $p_{Tcut}$, we normalize to the cross section
calculated as described above.  This yields most of the signal cross
section ($Wbq$ in the fiducial region).  If the $p_T$ of the $\bar b$
antiquark is {\it above} $p_{Tcut}$, we simply use the cross section
obtained from Fig.~\ref{Feynman}(b).  This yields the final state
$Wb\bar bq$, which we reject if all three jets are in the fiducial
region, but which contributes to the signal if one jet is missed, one
and only one of the two remaining jets is $b$-tagged and it, together
with the $W$ boson, reconstructs to the top-quark mass (within some
resolution).  This strategy avoids the occurrence of powers of
$\alpha_s\ln (m_t^2/m_b^2)$ at higher orders in perturbation theory,
which would degrade the accuracy of the calculation.

\subsection{Theoretical uncertainty}

\indent\indent In Ref.~\cite{SSW}, we studied the uncertainty in the
next-to-leading-order total cross section for single-top-quark
production via $W$-gluon fusion by varying the scale in the $b$-quark
distribution function and the strong coupling.  This indicated an
uncertainty in the total cross section of $\pm 5\%$, not including the
uncertainty in the parton distribution functions and the top-quark
mass.  However, to obtain the cross section with the $p_T$ of the
$\bar b$ antiquark below $p_{Tcut}$, we need to subtract from the
total cross section the cross section with the $p_T$ of the $\bar b$
antiquark above $p_{Tcut}$, as discussed above.  Since the latter is a
tree-level calculation, its scale dependence is relatively large.  We
use the scale $\mu^2 = p_T^2 + m_b^2$ in the gluon distribution
function and the strong coupling, and find a $\pm 30\%$ uncertainty in
this cross section at the Tevatron ($\pm 15\%$ at the LHC) by varying
$\mu$ between one half and twice this value.  We add in quadrature the
absolute uncertainty in the total cross section and the cross section
with the $p_T$ of the $\bar b$ antiquark above $p_{Tcut}$.  This
yields a relative uncertainty in the cross section with the $p_T$ of
the $\bar b$ antiquark below $p_{Tcut}$ of about $\pm 10\%$ at both
the Tevatron and the LHC.\footnote{Although the uncertainty in the
cross section with the $p_T$ of the $\bar b$ antiquark above
$p_{Tcut}$ is half as large at the LHC compared with the Tevatron, the
cross section itself is a larger fraction of the total cross section
at the LHC (see Table~\ref{sigma}).  This is why the uncertainty in
the cross section with the $p_T$ of the $\bar b$ antiquark below
$p_{Tcut}$ is comparable at the Tevatron and the LHC.}  This is
reflected in the uncertainty in the numbers in the last column of
Table~\ref{sigma}.  To reduce this uncertainty would require the
resummation of the large logarithms $\alpha_s^n \ln^n
[m_t^2/(p_{Tcut}^2+m_b^2)]/n!$ which appear in the calculation of the
cross section with the $p_T$ of the $\bar b$ antiquark above
$p_{Tcut}$.

Another source of uncertainty in the cross section is the uncertainty
in the parton distribution functions, especially the gluon
distribution function.  This uncertainty has recently been studied in
Ref.~\cite{HKLOOST}, and it appears to be less than $\pm 10\%$ at both
the Tevatron and the LHC.  This is comparable to the uncertainty
stemming from the scale variation described above.  That study
indicates that the uncertainty in the parton distribution functions
could potentially be pushed below $\pm 10\%$.

The uncertainty in the top-quark mass also leads to an uncertainty in
the cross section \cite{SSW}.  The present uncertainty in the
top-quark mass of $\pm 5.2$ GeV \cite{TOPMASS} corresponds to an
uncertainty in the cross section of $\pm 9\%$ at the Tevatron.
Anticipating an uncertainty of $\pm 3$ GeV from Run II at the Tevatron
\cite{tev2000} corresponds to an uncertainty in the cross section of
$\pm 5\%$, much less than the uncertainty from the scale variation and
the parton distribution functions.  The uncertainty in the top-quark
mass at the Tevatron and the LHC will ultimately reach $\pm 2$ GeV or
less \cite{tev2000}, corresponding to an uncertainty in the cross
section of $\pm 3\%$ at the Tevatron, and $\pm 2\%$ at the LHC.

Combining all theoretical uncertainties in quadrature, we estimate a
theoretical uncertainty of about $\pm 15\%$ in the cross section at
the Tevatron and the LHC, assuming an uncertainty in $m_t$ of $\pm 3$
GeV or less.

\section{Calculation}

\indent\indent The top-quark mass is taken to be 175 GeV
\cite{TOPMASS}. We optimize our study for the dominant
single-top-quark production mechanism, $W$-gluon fusion.  The final
state, $Wb\bar bj$, consists of a recoiling light-quark jet from the
production of the $t$-channel $W$ boson, a $\bar b$ antiquark from the
splitting of the initial gluon, and the decay products of the top
quark.  As discussed in the previous section, the large $t\bar t$
background requires that we use $Wbj$ as our signal, {\it i.e.}, we
reject events in which the $\bar b$ antiquark is detected above some
$p_{Tcut}$.  Thus our signal is a leptonically-decaying $W$ boson (to
reduce QCD backgrounds) plus two jets, with one and only one $b$ tag.
In addition to the $t\bar t$ background, the other principal
backgrounds are $Wb\bar b$ and $Wjj$ (with one jet mistagged), as well
as $Wc\bar c$ and $Wcj$ (with one $c$ quark mistagged).  The
background $WZ$, with $Z\to b\bar b$, is small and can be
neglected.\footnote{In contrast, $WZ$ with $Z\to b\bar b$ is an
important background to $WH$ with $H\to b\bar b$, because $M_Z$ is
near $m_H$ in the Higgs mass range of interest \cite{SMW}.}  Requiring
one and only one $b$ tag helps reduce the $t\bar t \to WWb\bar b$,
$Wb\bar b$, and $Wc\bar c$ backgrounds, while maintaining almost all
of the signal.

The signal and backgrounds for single-top-quark production are
calculated using tree-level matrix elements generated by MadGraph
\cite{MadGraph}.  The normalization of the $W$-gluon-fusion cross
section is determined as described in Section 2.  The $s$-channel
process is normalized to the next-to-leading-order cross section
\cite{SmithW}.  The $t \bar t$ cross section is normalized to the
next-to-leading-order result~\cite{NDE,BKVS},\footnote{We use the
central values given in the last paper of Ref.~\cite{CMNT}.}  not
including soft-gluon resummation \cite{CMNT,LSV,BC}.  The $W b \bar
b$, $Wc\bar c$, $Wcj$, and $Wjj$ cross sections are calculated at
leading order using the CTEQ4L~\cite{CTEQ4} parton distribution
functions with the renormalization and factorization scales chosen to
be $\mu^2 = \hat s$.  Since these cross sections will be measured in
the invariant-mass regions away from the top-quark mass, theoretical
uncertainties in the normalization of these backgrounds will not limit
the accuracy of the measurement of the signal cross section.  The $gb
\to Wt$ cross section is also calculated at leading order using CTEQ4L
and $\mu^2 = \hat s$.

We smear the jet energies with a Gaussian function of width $\Delta
E_j/E_j = 0.80/\sqrt {E_j} \oplus 0.05$ (added in quadrature) to
simulate the resolution of the hadron calorimeter. The momenta of
overlapping jets ($\Delta R_{jj}<0.7$) are added and the resulting
momentum is associated with a single jet.  We do not smear the lepton
energy, since this is a small effect compared with the smearing of the
jet energies. The lepton must be separated from the jets ($\Delta
R_{j\ell}>0.7$) or it is considered missed.  The two solutions for the
neutrino momentum which satisfy the missing-$p_T$ and $W$-mass
constraints are reconstructed and the solution with the smallest
magnitude of rapidity is chosen. This reconstructed event must pass
the cuts listed in Table~\ref{acceptance} used to simulate the
acceptance of the detector.  The rapidity and $p_T$ coverage are
chosen to simulate a generic detector.  Most of the jets are central
at the Tevatron, so it is only necessary to have jet coverage to
$|\eta_j|<2.5$, while the jets are distributed over a wider range of
rapidities at the LHC, necessitating converage to
$|\eta_j|<4$. Experimental results will by modified depending on
actual detector capabilities.

We assume a $b$-tagging efficiency of $60\%$ ($50\%$ for Run I) with a
mistag rate of $15\%$ for charm quarks and $0.5\%$ for light quarks at
both the Tevatron \cite{tev2000,Yao} and the LHC \cite{ATLAS}.  As we
shall see, the large charm background suggests it may be advantageous
to employ a strategy to reject charm (and light-quark) jets.  We
quantify the usefulness of increased charm and light-quark rejection
in the search for single-top-quark production in the Tevatron Run I
data.

\begin{table}
\begin{center}
\caption{Cuts used to simulate the acceptance of the detector. The
rapidity coverage for jets is taken to be $|\eta_j|<2.5$ at the
Tevatron and $|\eta_j|<4$ at the LHC.  The rapidity coverage for $b$
tagging is taken to be $|\eta_b|<1$ at Tevatron Run I, and
$|\eta_b|<2$ at Tevatron Run II and beyond, as well as at the LHC.
The $p_{T\ell}$ threshold is greater for charged leptons which are
used as triggers (in parentheses).
\label{acceptance}} 
\bigskip 
\begin{tabular}{ll}
$|\eta_b|<2$ (1) & $p_{Tb}>20$ GeV \\
$|\eta_\ell|<2.5$ & $p_{T\ell}>10$ GeV (20 GeV)\\
$|\eta_j|<2.5$ (4) & $p_{Tj}>20$ GeV \\
$|\Delta R_{jj}|>0.7 $ & $|\Delta R_{j\ell}|>0.7$ \\
${\not \!p}_{T}>20$ GeV & \\
\end{tabular} \end{center} 
\end{table}

The $p_T$ spectrum of the $\bar b$ antiquark, the $b$ quark from the
top decay, and the light-quark jet from the emission of the
$t$-channel $W$ boson, from single-top-quark production via $W$-gluon
fusion, are shown in Fig.~\ref{b_pt}.  The $b$-quark $p_T$ spectrum
peaks at about 60 GeV, and the light-quark jet has a broad $p_T$
spectrum, while the $\bar b$ antiquark is produced mostly at low
$p_T$.  Hence the majority of our signal comes from tagging the $b$
quark, with the light quark providing the second jet.\footnote{After
all cuts, this accounts for about $94\%$ of the signal at both the
Tevatron and the LHC.}  However, we include in our signal {\em any}
final state with two and only two jets with $p_T>20$ GeV, with one and
only one $b$ tag.

\section{Results}

\indent\indent Our results are summarized in Table~\ref{results}.  The
first column shows the total cross section times the branching ratio
(2/9) for the top quark to decay semileptonically (not including the
$\tau$ lepton, which is treated as a jet). The signal cross section
includes both $t$ and $\bar t$ production.  Similarly, the $Wb\bar b$,
$Wc\bar c$, $Wcj$, and $Wjj$ backgrounds account for both $W^+$ and
$W^-$ production times the branching ratio 2/9. The $t \bar t$
background is multiplied by the branching ratio 4/9 to include the
possibility that either the $t$ or the ${\bar t}$ decays
semileptonically (a $t\bar t$ event can be a background to either
single $t$ or single $\bar t$ production).

The second column in Table~\ref{results} shows the cross section for
events which pass the detector acceptance.  These events have one and
only one $b$-tagged jet, and at least one other jet.  The numbers in
parentheses are the cross sections for events which have a
reconstructed $b\ell\nu$ invariant mass within $\pm 20$ GeV of the
top-quark mass (to account for jet resolution).  About $70\%$ of the
single-top-quark events from $W$-gluon fusion survive this cut at the
Tevatron ($60\%$ at the LHC), while only about $40\%$ of the $t \bar
t$ events survive, and only $20\%$ of the $W b \bar b$, $Wc\bar c$,
$Wcj$, and $Wjj$ events survive (at both machines).  The low
acceptance for these last four backgrounds is easily understood since
there is no kinematic preference towards the top-quark mass. The $t
\bar t$ acceptance is only about 40\% because one half of the time the
tagged $b$ quark is associated with the other top-quark in the event.

\begin{table}
\caption{Cross sections (fb) for single-top-quark production and a
variety of background processes at the Tevatron and the LHC.  The
$W$-gluon-fusion signal is denoted by $t\bar b j$, and the $s$-channel
process by $t\bar b$. The first column is the total cross section for
$t$ + $\bar t$ production times the branching ratio (2/9) of the $W$
boson to $e,\mu$. The $t \bar t$ background is multiplied by a
branching ratio of 4/9 to account for either the $t$ or the $\bar t$
decaying semileptonically. The second column adds the cuts listed in
Table~\ref{acceptance} to simulate the acceptance of the detector, and
also includes a $b$-tagging efficiency of $60\%$ ($50\%$ for Run I)
with a mistag rate of $15\%$ for charm and $0.5\%$ for light-quark
jets.  Listed in parentheses is the cross section for events in which
the reconstructed $b\ell \nu$ invariant mass is within $\pm 20$~GeV of
the top-quark mass.  The third column includes a jet veto $p_T <
$20~GeV to reduce the $t\bar t$ background.\label{results}}
\medskip
\begin{center}
\begin{tabular}{ccccc}
&Tevatron&1.8 TeV $p\bar p$&\\
&Total$\times$ BR &Detector (peak)&Veto (peak)\medskip\\
$t\bar bj$ & 378  &  61      (41) &  46  (33)  \\
$t\bar b$  & 162  &  36      (18) &  36  (18)  \\
$Wt     $  &  16  &  6.1    (3.1) & 1.4 (0.8)  \medskip\\
$Wb\bar b$ & 6500 & 106      (20) & 106  (20)  \\
$Wc\bar c$ & ---  & 44        (8) &  44   (8)  \\
$Wcj$	   & ---  & 136      (24) & 136  (24)  \\
$Wjj$      & ---  & 127      (25) & 127  (25)  \\
$t\bar t$  & 2160 & 551     (240) &  47  (13)  \\
\\
&Tevatron&2 TeV $p\bar p$&\\
&Total$\times$ BR &Detector (peak)&Veto (peak)\medskip\\
$t\bar bj$ & 542  & 133      (90) & 107  (76)  \\
$t\bar b$  & 196  &  48      (24) &  48  (24)  \\
$Wt     $  &  26  &  16     (8.4) & 3.8 (2.1)  \medskip\\
$Wb\bar b$ & 7420 & 146      (28) & 146  (28)  \\
$Wc\bar c$ & ---  & 74       (14) &  74  (14)  \\
$Wcj$	   & ---  & 274      (53) & 274  (53)  \\
$Wjj$      & ---  & 257      (54) & 257  (54)  \\
$t\bar t$  & 2980 & 838     (364) &  80  (24)  \\
\\
&LHC&14 TeV $p p$&\\
&Total$\times$ BR &Detector (peak)&Veto (peak)\medskip\\
$t\bar bj$ & 54400 & 12500  (7510)& 8930 (6110)\\
$t\bar b$  & 2270  &  470   (229) &  470 (229) \\
$Wt     $  & 13700 &  7510  (3610)& 1650 (820) \medskip\\
$Wb\bar b$ & 70700 & 1140    (230)& 1140 (230) \\
$Wc\bar c$ & ---   &  750    (150)&  750 (150) \\
$Wcj$	   & ---   & 24200  (5070)&24200 (5070)\\
$Wjj$      & ---   & 7000   (1460)& 7000 (1460)\\
$t\bar t$  & 357000& 95600 (40700)& 9040 (2770)\\
\\
\end{tabular}
\end{center}
\end{table}

It is evident from Table~\ref{results} that the largest background is
$t\bar t \to W^+W^-b\bar b$, and it is much larger than the
signal. This background is particularly worrisome because it produces
a peak in the $b\ell\nu$ invariant-mass spectrum at the top-quark
mass, just as does the signal.\footnote{In contrast, the $t\bar t$
background is not as problematic for the process $WH$ with $H\to b\bar
b$, because it does not produce a peak in the $b\bar b$ invariant mass
near the Higgs mass \cite{SMW}.}  Hence it is important to apply
additional cuts to reduce this background.  Since this background has
an additional $W$ boson in the final state, we reject events which
have an additional charged lepton\footnote{The $\tau$ lepton is
treated as a jet.} with $p_{T\ell}>10$ GeV, or an additional jet with
$p_{Tj} > 20$ GeV and $|\eta_j| < 2.5$ at the
Tevatron\footnote{Increasing the jet rapidity coverage to $|\eta_j| <
4$ at the Tevatron does not decrease this background significantly.}
($|\eta_j| < 4$ at the LHC). This reduces the $t\bar t$ background by
a factor of 15 in the peak region at both the Tevatron and the LHC,
while reducing the signal by only a modest amount, since the signal
rarely has a third jet with $p_{Tj}> 20$ GeV.  This ``veto''
\cite{BCHZ} yields the signal and background cross sections listed in
the third column of Table~\ref{results}.

\begin{figure}[tbp]
\begin{center}
\input{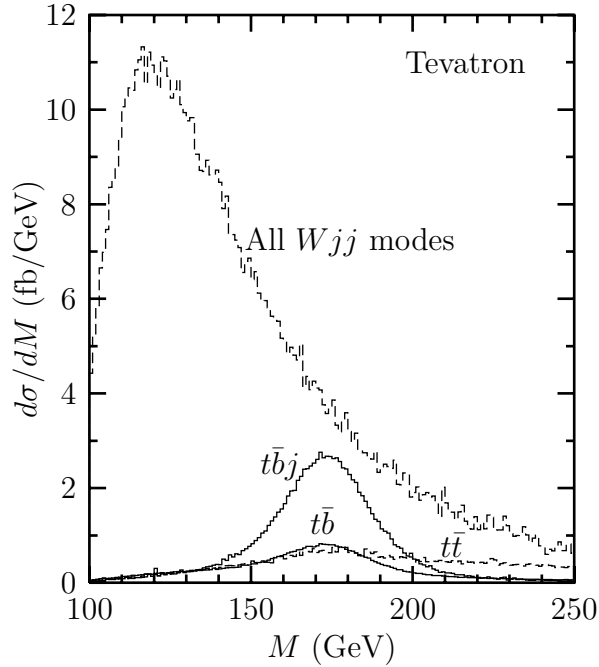}        
\end{center}
\caption{The $b\ell\nu$ invariant-mass ($M$) distribution for
single-top-quark production and backgrounds at the Tevatron ($\sqrt
S=2$ TeV) with single $b$ tagging.  The $W$-gluon-fusion signal is
denoted by $t\bar b j$, and the $s$-channel process by $t\bar b$ (the
$Wt$ process is negligibly small).  The $Wjj$ background includes
$Wb\bar b$, $Wc\bar c$, $Wcj$, and $Wjj$.  The $t\bar t$ background is
shown separately.\label{mtopt2}}
\end{figure}

\begin{figure}[tbp]
\begin{center}
\input{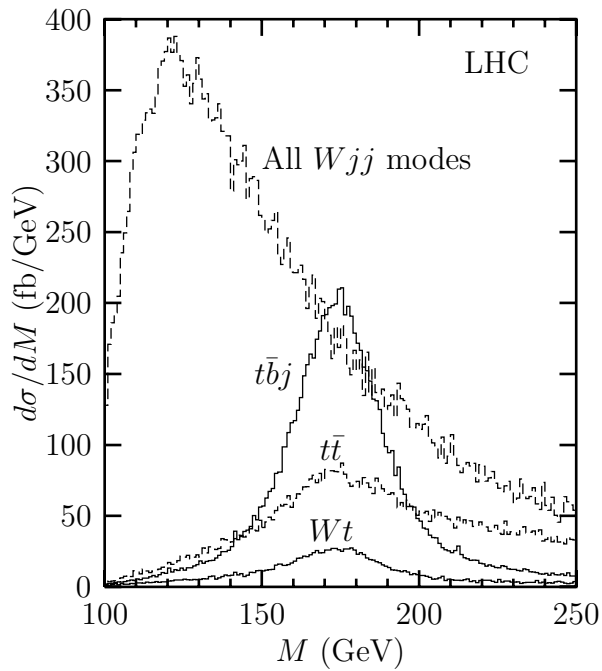}    
\end{center}
\caption{The $b\ell\nu$ invariant-mass ($M$) distribution for
single-top-quark production and backgrounds at the LHC with single $b$ 
tagging.  The $W$-gluon-fusion signal is denoted by $t\bar b j$,
and the $Wt$ process is also shown (the $s$-channel process is 
negligibly small).  The $Wjj$ background includes $Wb\bar b$, 
$Wc\bar c$, $Wcj$, and $Wjj$.  The $t\bar t$ background is shown separately.
\label{mtopl}}
\end{figure}

We show the $b\ell\nu$ invariant-mass distribution for
single-top-quark production and the various backgrounds at the
Tevatron ($\sqrt S = 2$ TeV) in Fig.~\ref{mtopt2}, and at the LHC in
Fig.~\ref{mtopl}.  The $W$-gluon-fusion process is prominent at both
the Tevatron and the LHC,\footnote{One of the factors contributing to
the prominence of the signal, in comparison with the study in
Ref.~\cite{tev2000}, is that a poorer jet energy resolution was
assumed in that study.}  but the backgrounds are non-negligible.  The
$t\bar t$ background has been reduced to an acceptable level, but it
is still significant, and because it has the same shape as the signal
it will be necessary to know the normalization of this background
independently.  This could be achieved by measuring the $t\bar t$
cross section using the full $W^+W^-b\bar b$ final state, and then
calculating its contribution to the $Wbj$ background, as we have done.
Since we desire to separate single-top-quark production via $W$-gluon
fusion from the $s$-channel process, it will also be necessary to
measure the latter and subtract it from the signal.  This can be
achieved by double $b$ tagging \cite{SW}, as discussed in Section 5.
This is unnecessary at the LHC, where the $s$-channel process is
negligible.

The remaining backgrounds --- $Wb\bar b$, $Wc\bar c$, $Wcj$, and $Wjj$
--- all yield continuous spectra, and can therefore be calibrated by
measuring them in the invariant-mass regions away from the peak
region.  These backgrounds are significant and comparable to each
other at the Tevatron, but only $Wcj$ is significant at the LHC. It
may be desirable to reject more strongly events in which a charm or
light quark fakes a $b$ jet, at both the Tevatron and the LHC.  The
VXD3 vertex detector in SLD has achieved a $b$-tagging efficiency of
$50\%$, with a mistag rate of only $1.24\%$ from charm and $0.07\%$
from light quarks \cite{VXD2,VXD3}.

The statistics for discovering a signal are different from those for
measuring its cross section.  To claim a discovery, one needs to
demonstrate that the signal is not consistent with a fluctuation in
the background.  The discovery significance is therefore governed by
the number of signal events divided by the square root of the number
of background events, $S/\sqrt{B}$.  On the other hand, the accuracy
with which a cross section can be measured is limited by the
fluctuation in the total number of expected events in the signal
region, $S+B$.  Thus the fractional uncertainty in the measured cross
section is $\sqrt {S+B}/S$.

We list in Table~\ref{signif} the statistical significance of the
single-top-quark signal, and the statistical uncertainty in the
measured cross section, at the Tevatron and the LHC.  All three
single-top-quark production processes have been regarded as part of
the signal in determining these numbers, although $W$-gluon fusion is
dominant.\footnote{The significance and statistical sensitivity for
single-top-quark production via $W$-gluon fusion alone are somewhat
less, because the $s$-channel process and the $Wt$ process must then
be considered as backgrounds.}  We see that there is not sufficient
data to discover single-top-quark production in the Tevatron Run I
data, and the cross section can be measured only crudely.  As
mentioned above, it may be possible to increase the rejection of charm
and light-quark jets.  However, even if it were possible to achieve
100\% rejection while maintaining the 50\% $b$-tagging efficiency the
significance of the signal would be only $3\sigma$, not enough for
discovery, but perhaps enough for ``evidence'' of single-top-quark
production in Run I.

\begin{table}
\begin{center}
\caption{Statistical significance of the signal ($S/\sqrt B$) and
accuracy of the measured cross section ($\sqrt {S+B}/S$) for
single-top-quark production via $W$-gluon fusion at the Tevatron and
the LHC.  Also given is the accuracy of the extracted value of
$V_{tb}$, assuming a $\pm 15\%$ uncertainty in the theoretical cross
section.\label{signif}}
\bigskip 
\begin{tabular}{cccc}
& $S/\sqrt B$ & $\sqrt{S+B}/S$  & $\Delta V_{tb}/V_{tb}$ \\
\\
1.8 TeV $p\bar p$ (110 pb$^{-1}$) & 1.8 & 69\% & 35\% \\
2 TeV $p\bar p$ (2 fb$^{-1}$)   & 11  & 12\% & 10\% \\
2 TeV $p\bar p$ (30 fb$^{-1}$)  & 43  & 3.0\% & 7.6\% \\
14 TeV $pp$ (1 fb$^{-1}$)         & 73  & 1.8\% & 7.6\% \\
\end{tabular} \end{center} 
\end{table}

Single-top-quark production should be discovered ($5\sigma$) at Run II
with about 500 pb$^{-1}$ of integrated luminosity, and the cross section will
ultimately be measured to an accuracy of $\pm 12\%$ with 2 fb$^{-1}$
of integrated luminosity.  This is comparable to the theoretical accuracy, which we
estimated to be $\pm 15\%$ in Section 2.1.  Combined in quadrature,
and neglecting any systematic uncertainties, we conclude that $V_{tb}$
can be measured to an accuracy of $\pm 10\%$ in Run II at the Tevatron
(assuming $V_{tb} \approx 1$).

Additional running at the Tevatron will reduce the statistical
uncertainty further.  An integrated luminosity of 30 fb$^{-1}$ yields
a statistical uncertainty in the cross section of only $\pm 3\%$.
Together with the $\pm 15\%$ theoretical uncertainty, this yields an
uncertainty in $V_{tb}$ of $\pm 7.6\%$.  In order to maximally benefit
from the reduced statistical uncertainty, it is necessary to reduce
the theoretical and systematic uncertainties to a level comparable to
the statistical uncertainty.  The study of Ref.~\cite{HKLOOST}
suggests that the uncertainty in the quark-gluon luminosity can be
reduced below $\pm 10\%$.  The scale uncertainty in the cross section
of $\pm 10\%$ requires additional theoretical work to reduce, as
discussed in Section 2.1.  We have not attempted to estimate the
experimental systematic uncertainties.

The statistical uncertainty at the LHC is only 1.8\% with just 1
fb$^{-1}$ of integrated luminosity.  The accuracy with which $V_{tb}$
can be measured will therefore be limited entirely by the theoretical
and systematic uncertainties.  It is a challenge to reduce these to a
level such that one can benefit from the tremendous statistical
sensitivity of the LHC.

\subsection{Forward jet tag}

The emission of the virtual $t$-channel $W$ boson in single-top-quark
production via $W$-gluon fusion [Fig.~\ref{Feynman}(b)] results in a
high-rapidity jet in the final state \cite{Y,EP,HBB}.  The same
phenomenon occurs for Higgs production via $WW$ fusion, and the
tagging of this forward jet has been advocated to isolate the signal
from the background for a heavy Higgs boson \cite{BCHZ,KS,BHP,BG}.  In
this section we investigate whether this feature of $W$-gluon fusion
can be used to increase the statistical significance ($S/\sqrt B$) of
our signal.

\begin{figure}[tbp]
\begin{center}
\input{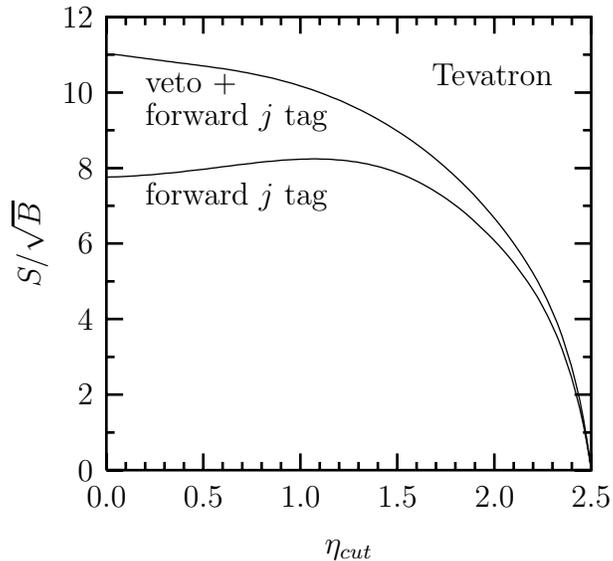}        
\end{center}
\caption{Significance of the single-top-quark signal in Run II at the 
Tevatron ($\sqrt S=2$ TeV, 2 fb$^{-1}$) versus the minimum rapidity
of the non-$b$-tagged jet in the signal.  Curves are shown with and without
the jet veto imposed.\label{fort2}}
\end{figure}

We perform the same analysis as above, but we demand that the
non-$b$-tagged jet have a rapidity whose magnitude is greater than
$\eta_{cut}$.  Before imposing the jet veto, we find that the
significance rises slightly as $\eta_{cut}$ is increased from zero,
and then eventually decreases.  This is shown in Fig.~\ref{fort2} at
the Tevatron ($\sqrt S=2$ TeV), and the result is similar at the LHC.
However, after imposing the jet veto, we find that the significance
decreases as $\eta_{cut}$ is increased from zero, and is always
greater than the signficance without the veto.  This is also shown in
Fig.~\ref{fort2}.  Thus this simple forward jet tag does not increase
the significance of the signal.  However, this forward jet is a
characteristic of single-top-quark production via $W$-gluon fusion,
and its observation would build confidence that one has observed this
process.

\section{Comparison with $s$-channel single-top-quark production}

\indent\indent The $s$-channel production of single top quarks, shown
in Fig.~\ref{Feynman}(a), also provides a means to measure the CKM
matrix element $V_{tb}$ \cite{CP,SW}.  Furthermore, the $s$-channel
and $t$-channel ($W$-gluon fusion) processes generally have different
dependence on new physics, so it is worthwhile to measure them
separately \cite{CY,CMY,ABES,DZ,LOY,S,LCHZX,DYYZ,TY}.  In this section
we calculate the sensitivity of the Tevatron to $V_{tb}$ via the
$s$-channel process, and compare it with the results for $W$-gluon
fusion presented in the previous section.

The final state in $s$-channel production of single top quarks is
$Wb\bar b$.  It can be separated from single-top-quark production via
$W$-gluon fusion by double $b$ tagging, since $W$-gluon fusion usually
produces only one $b$ jet with $p_T > 20$ GeV.  The $s$-channel
process has a smaller cross section than $W$-gluon fusion, and the
efficiency of double $b$-tagging is less than that of single $b$
tagging, so the statistical sensitivity is less for the $s$-channel
process.  However, this is compensated by the smaller theoretical
uncertainty in the cross section \cite{SW,SmithW}.

The analysis of signal and backgrounds follows closely that of
$W$-gluon fusion.  However, because we now demand two $b$ tags, the
backgrounds are generally smaller, with the exception of $Wb\bar b$
and $t\bar t \to WWb\bar b$, which also contain two $b$ jets in the
final state.  To select the correct $b$ jet to associate with the top
quark, we use the fact that in $q\bar q \to t\bar b$ the $b$ quark
from top decay tends to go in the proton direction in $p\bar p$
collisions \cite{CP,SW}.  As in the $W$-gluon-fusion analysis, we
choose the solution for the neutrino momentum which has the smallest
magnitude of rapidity.\footnote{Our analysis is essentially the same
as Ref.~\cite{SW}, with the exception that we do not make the cut
$M(b\bar b)> 110$ GeV, which was made to suppress the $WZ \to Wb\bar
b$ background.  We have found this cut to be unnecessary, as this
background is modest in the signal region, $M(b\ell\nu) \approx m_t$,
as evidenced by the results in Table~\ref{results2}.}

\begin{figure}[tbp]
\begin{center}
\input{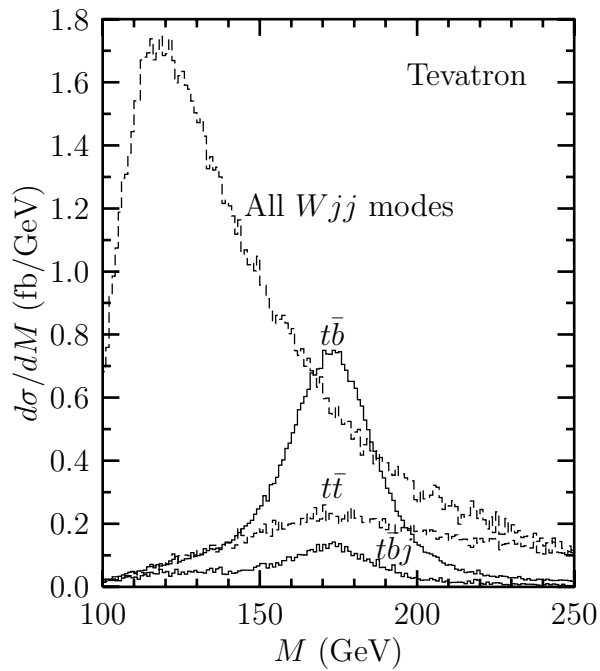}        
\end{center}
\caption{The $b\ell\nu$ invariant-mass ($M$) distribution for
single-top-quark production and backgrounds at the Tevatron ($\sqrt
S=2$ TeV) with double $b$ tagging.  The $s$-channel signal is denoted
by $t\bar b$, and the $W$-gluon-fusion process by $t\bar bj$ (the $Wt$
process is negligibly small).  The $Wjj$ background is dominated by
$Wb\bar b$.  The $t\bar t$ background is shown
separately.\label{smtopt2}} 
\end{figure}

The results are presented in Table~\ref{results2}.  As in the
$W$-gluon fusion analysis, a jet veto is necessary to suppress the
enormous $t\bar t$ background.  The resulting $b\ell\nu$
invariant-mass distribution is show in Fig.~\ref{smtopt2} at the
Tevatron ($\sqrt S = 2$ TeV) and in Fig.~\ref{smtopl} at the LHC.  The
signal is prominent at the Tevatron, although the backgrounds are
non-negligible.  The situation is less promising at the LHC, due to
the large $t\bar t$ background, which has the same shape as the
signal.  Furthermore, the majority of signal events come from
$W$-gluon fusion, not from the $s$-channel process, so the
double-$b$-tag strategy does not succeed in isolating the $s$-channel
process at the LHC.  We henceforth concentrate our analysis on the
Tevatron results.

\begin{figure}[tbp]
\begin{center}
\input{smtopl}    
\end{center}
\caption{The $b\ell\nu$ invariant-mass ($M$) distribution for
single-top-quark production and backgrounds at the LHC with double $b$
tagging.  The $s$-channel signal is denoted by $t\bar b$, and the
$W$-gluon-fusion process by $t\bar bj$ (the $Wt$ process is negligibly
small).  The $Wjj$ background is dominated by $Wb\bar b$.  The $t\bar
t$ background is shown separately.\label{smtopl}}
\end{figure}

\begin{table}
\caption[fake]{Cross sections (fb) for single-top-quark production
and a variety of background processes at the Tevatron and the
LHC.  The $s$-channel signal is denoted by $t\bar b$, and the 
$W$-gluon-fusion process is denoted by $t\bar b j$. The analysis is
as described in the caption of Table~\ref{results}, except we have 
required two $b$ tags instead of one and only one $b$ tag.
\label{results2}}
\medskip
\begin{center}
\begin{tabular}{ccccc}
&Tevatron&1.8 TeV $p\bar p$&\\
&Total$\times$ BR &Detector (peak)&Veto (peak)\medskip\\
$t\bar b$  & 162  &  8.9    (5.5) & 8.9  (5.5) \\
$t\bar bj$ & 378  &  4.4    (2.3) & 1.5  (0.7) \\
$Wt$       & 16   & 0.04   (0.01) & 0.03 (0.01) \medskip\\
$Wb\bar b$ & 6500 & 29      (5.1) & 29   (5.1)  \\
$WZ$	   & 58   & 2.5	    (0.6) & 2.5  (0.6) \\
$Wc\bar c$ & ---  & 2.6     (0.5) & 2.6  (0.5) \\
$Wcj$	   & ---  & 0.35     (0.06) & 0.35  (0.06) \\
$Wjj$      & ---  & 0.28   (0.05) & 0.28 (0.05) \\
$t\bar t$  & 2160 & 136     (61) &  8.4  (2.6)   \\
\\
&Tevatron&2 TeV $p\bar p$&\\
&Total$\times$ BR &Detector (peak)&Veto (peak)\medskip\\
$t\bar b$  & 196  &  32      (21) &  32  (21)  \\
$t\bar bj$ & 542  &  21      (11) &   7  (4)  \\
$Wt$       & 26   & 0.1    (0.04) & 0.1 (0.04) \medskip\\
$Wb\bar b$ & 7420 &  97      (18) &  97  (18)  \\
$WZ$	   & 71   & 10	    (2.2) &  10  (2.2) \\
$Wc\bar c$ & ---  &  6.3    (1.2) & 6.3  (1.2)  \\
$Wcj$	   & ---  &  1.2    (0.2) & 1.2  (0.2)  \\
$Wjj$      & ---  & 0.6     (0.1) & 0.6 (0.1)  \\
$t\bar t$  & 2980 & 496     (223) & 28   (8)  \\
\\
&LHC&14 TeV $p p$&\\
&Total$\times$ BR &Detector (peak)&Veto (peak)\medskip\\
$t\bar b$  & 2270  &  209    (103)&  209 (103) \\
$t\bar bj$ & 54400 & 2055    (932)&  492 (221) \\
$Wt$	   & 13700 &   44     (15)&   41  (14) \medskip\\
$Wb\bar b$ & 70700 &  544    (112)&  544 (112) \\
$WZ$	   & 880   &   50     (14)&   50  (14) \\
$Wc\bar c$ & ---   &   51     (10)&   51  (10) \\
$Wcj$	   & ---   &   83     (17)&   83  (17) \\
$Wjj$      & ---   &   13      (3)&   13   (3) \\
$t\bar t$  & 357000& 44800 (19500)& 2780 (838) \\
\\
\end{tabular}
\end{center}
\end{table}

We list in Table~\ref{signif2} the discovery significance ($S/\sqrt
B$) and the statistical accuracy ($\sqrt{S+B}/S$) for $s$-channel
production of single top quarks.  As with $W$-gluon fusion, there is
not enough data in Run I for discovery.

\begin{table}
\begin{center}
\caption{Statistical significance of the signal ($S/\sqrt B$) and
accuracy of the measured cross section ($\sqrt {S+B}/S$) for
single-top-quark production via the $s$-channel process at the
Tevatron.  Also given is the accuracy of the extracted value of
$V_{tb}$, assuming a $\pm 6\%$ uncertainty in the theoretical cross
section.  We include an uncertainty in the theoretical cross section,
due to the uncertainty in the top-quark mass, of $\pm 15\%$ at Run I,
$\pm 7.5\%$ at Run II, and $\pm 5\%$ beyond Run II at the
Tevatron. \label{signif2}}
\bigskip 
\begin{tabular}{cccc}
& $S/\sqrt B$ & $\sqrt{S+B}/S$  & $\Delta V_{tb}/V_{tb}$ \\
\\
1.8 TeV $p\bar p$ (110 pb$^{-1}$) & 0.7 & 190\% & 95\% \\
2 TeV $p\bar p$ (2 fb$^{-1}$)     & 6.5 & 21\% & 12\% \\
2 TeV $p\bar p$ (30 fb$^{-1}$)    & 25  & 5.4\% & 4.7\% \\
\end{tabular} \end{center} 
\end{table}

Discovery ($5\sigma$) will occur in Run II after approximately 1
fb$^{-1}$ of integrated luminosity has been collected, and the cross
section will ultimately be measured to $\pm 21\%$ with 2 fb$^{-1}$ of
data.  The theoretical uncertainty in the cross section is estimated
to be about $\pm 6\%$ \cite{SmithW}, plus an additional uncertainty
from the uncertainty in the top-quark mass.  Assuming the mass is
measured to $\pm 3$ GeV in Run II \cite{tev2000}, this adds an
uncertainty in the cross section of $\pm 7.5\%$ \cite{SmithW}.
Combining all three uncertainties in quadrature, one finds that
$V_{tb}$ will be measured to $\pm 12\%$ via the $s$-channel process in
Run II.  This is comparable to the accuracy achieved via $W$-gluon
fusion, which has a smaller statistical uncertainty but a larger
theoretical uncertainty.

The small theoretical uncertainty in the $s$-channel process becomes
increasingly relevant with greater integrated luminosity.  With 30
fb$^{-1}$ of integrated luminosity, the statistical uncertainty is
comparable to the theoretical uncertainty of $\pm 6\%$.\footnote{It is
likely that the theoretical uncertainty can be reduced below $\pm 6\%$
with additional work \cite{SmithW}.}  Assuming the top-quark mass can
be measured to $\pm 2$ GeV with this amount of data \cite{tev2000},
there is an additional $\pm 5\%$ uncertainty in the theoretical cross
section from the uncertainty in the top-quark mass.  Combining all
three uncertainties in quadrature, one finds that $V_{tb}$ can be
measured to $\pm 4.7\%$ with 30 fb$^{-1}$ of integrated luminosity at
the Tevatron.  This is better than the $\pm 7.6\%$ uncertainty
achieved via $W$-gluon fusion, due to its larger theoretical
uncertainty.

\section{Polarization}

\indent\indent Single-top-quark production proceeds via the weak
interaction, so the top quarks produced are highly polarized
\cite{CY,HBB}.  The optimal basis for the study of this polarization
was recently constructed in Ref.~\cite{MP}.  The top quark is 100\%
polarized, in the top-quark rest frame, along the direction of the $d$
quark (or $\bar d$ antiquark) in the event.  Since $W$-gluon fusion
proceeds via $ug \to dt\bar b$ about 77\% of the time at the Tevatron
($\sqrt S=2$ TeV), the $d$ quark is usually the light-quark jet. The
other 23\% of the events proceed via $\bar d g \to \bar u t\bar b$, in
which case the $\bar d$ quark is moving along one of the beam
directions.  However, since the light-(anti)quark ($\bar u$) jet tends
to move in the same direction as the $\bar d$ antiquark, the direction
of the light-quark jet is still a rather good basis to analyze the
spin in these events.  It is shown in Ref.~\cite{MP} that the top
quark has a net 96\% polarization along the direction of the
light-quark jet in single-top-quark production via $W$-gluon fusion at
the Tevatron ($\sqrt S=2$ TeV).

Since the top quark decays well before QCD interactions can flip its
spin, the polarization of the top quark is observable in the
distribution of its decay products \cite{BDKKZ}. The most sensitive
spin analyzer is the charged lepton in semileptonic decay, whose
partial width ($\Gamma$) has the angular distribution
\begin{equation}
\frac{1}{\Gamma}\frac{d\Gamma}{d\cos\theta} = \frac{1}{2}(1 + \cos\theta)
\label{cos}
\end{equation}
where $\theta$ is the angle, in the top-quark rest frame, between the
direction of the charged lepton and the spin of the top quark
\cite{J}.  Thus, in $W$-gluon-fusion events, the charged lepton from
top-quark decay has an angular distribution with respect to the
light-quark jet, in the top-quark rest frame, given approximately by
Eq.~(\ref{cos}).

\begin{figure}[tbp]
\begin{center}
\input{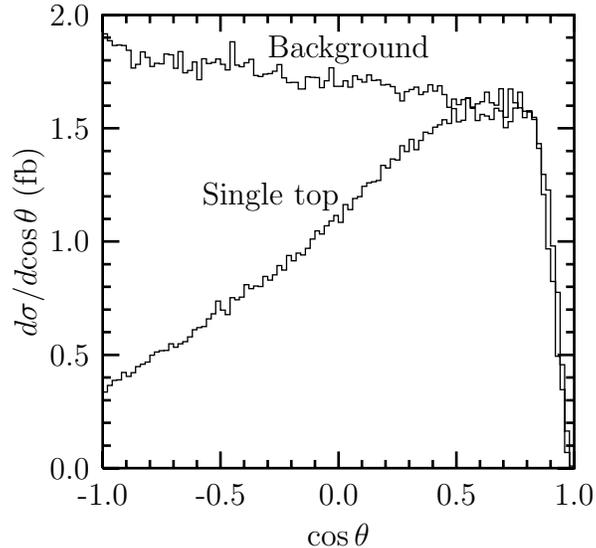} 
\end{center}
\caption{Angular distribution of the charged lepton in
single-top-quark events at the Tevatron ($\sqrt S=2$ TeV), with
respect to the non-$b$-tagged jet, in the top-quark rest frame.  Also
shown is the angular distribution of the sum of all background events.
The distributions correspond to the events in the last column of
Table~\ref{results}.\label{spin1}}
\end{figure}

We show in Fig.~\ref{spin1} the angular distribution of the charged
lepton with respect to the direction of the non-$b$-tagged jet, in the
top-quark rest frame, from single-top-quark events at the Tevatron
($\sqrt S=2$ TeV).  The expected angular distribution,
Eq.~(\ref{cos}), is approximately observed, but is degraded by the
cuts in Table~\ref{acceptance}, jet resolution, jet veto, and
reconstruction of the neutrino's momentum, as well as the contribution
from the $s$-channel process. The suppression at $\cos\theta \approx
1$ is due to the $\Delta R_{jl}>0.7$ cut between the charged lepton
and the jet.  The other curve is due to the sum of all the
backgrounds, and is nearly isotropic (except for the suppression at
$\cos\theta \approx 1$), despite the cuts imposed on the events.

A simple test to observe the top-quark polarization in $W$-gluon
fusion is to measure the asymmetry in the angular distribution of the
charged lepton, Fig.~\ref{spin1}.  Since the $\Delta R_{jl}>0.7$ cut
removes the small-angle region, we define an asymmetry between $-1 <
\cos\theta < 0.8$:
\begin{equation}
A \equiv \frac{\sigma(-1<\cos\theta<-0.1) - \sigma(-0.1<\cos\theta<0.8)}
	      {\sigma(-1<\cos\theta<-0.1) + \sigma(-0.1<\cos\theta<0.8)}  
\end{equation}
The signal in Fig.~\ref{spin1} has an asymmetry of $-38\%$.  An
unpolarized top quark would have zero asymmetry, so a nonzero
asymmetry measurement would constitute observation of the polarization
of the top-quark in $W$-gluon fusion events. Including the background,
the expected asymmetry measurment at the Tevatron is about $-14\%$.
Evidence for nonzero asymmetry (3$\sigma$) will be available in the
Run II data.  Observation at the 5$\sigma$ level takes approximately
500 signal events, which requires about 5 fb$^{-1}$ of integrated
luminosity.  The asymmetry is also evident at the LHC.

\section{Conclusions}

\indent\indent In this paper we have outlined a strategy to discover
single-top-quark production via $W$-gluon fusion and measure its cross
section at the Fermilab Tevatron and the CERN LHC.  The signal is
extracted by searching for a semileptonically-decaying top quark with
one $b$ tag, a non-$b$-tagged jet, and no additional jets or leptons.
We have also studied single-top-quark production via $q\bar q \to
t\bar b$, which can be separated from $W$-gluon fusion (at the
Tevatron) by requiring two $b$ tags.  These two single-top-quark
production processes provide independent measurements of the CKM
matrix element $V_{tb}$, and are generally influenced by new physics
in different ways.

Since the final state in single-top-quark production via $W$-gluon
fusion ($qg \to t\bar bq$) contains a $\bar b$ antiquark in addition
to the desired signal, we have calculated the cross section with the
$p_T$ of the $\bar b$ antiquark below 20 GeV.  This is a large
fraction of the total cross section, since the $\bar b$ antiquark
arises from the splitting of the initial gluon to a nearly-collinear
$b\bar b$ pair, and hence is usually at low $p_T$.  The calculation of
this cross section requires careful consideration of the collinear
region.  We obtained this cross section by subtracting the cross
section with $p_T > 20$ GeV from the next-to-leading-order total cross
section.  The resulting cross section, listed in Table~\ref{sigma},
has an uncertainty of about $\pm 10\%$ (estimated by varying the scale
in the gluon distribution function and the strong coupling), as well
as an additional uncertainty of $\pm 10\%$ from the parton
distribution functions, resulting in a total theoretical uncertainty
of $\pm 15\%$.

The accuracy with which $V_{tb}$ can be extracted from
single-top-quark production via $W$-gluon fusion is listed in
Table~\ref{signif}.  An accuracy of about $\pm 12\%$ should be
achieved in Run II at the Tevatron.  The accuracy saturates at $7.6\%$
(half the theoretical uncertainty) at both the Tevatron (30 fb$^{-1}$)
and the LHC.  Improving the accuracy therefore requires additional
work to reduce the theoretical uncertainty in the cross section.

We also considered single-top-quark production via $q\bar q \to t\bar
b$.  This process has a smaller theoretical uncertainty, but a larger
statistical uncertainty.  The accuracy with which $V_{tb}$ can be
extracted via this process is listed in Table~\ref{signif2}.  The
accuracy is only slightly worse than that of $W$-gluon fusion in Run
II at the Tevatron.  With additional integrated luminosity (30
fb$^{-1}$), an accuracy of $\pm 5\%$ can be achieved, somewhat better
than that of $W$-gluon fusion.  Single-top-quark production via $q\bar
q\to t\bar b$ is much more difficult to extract at the LHC due to the
large background from $t\bar t$ and $W$-gluon fusion.

We also considered the amount of data needed to detect the
polarization of the top quark in single-top-quark production via
$W$-gluon fusion.  We found that Run II at the Tevatron will produce
evidence for this effect.  Approximately 5 fb$^{-1}$ of integrated
luminosity is needed to establish the polarization at the 5$\sigma$
level.

Single-top-quark production via $W$-gluon fusion and $q\bar q\to t\bar
b$ represents an entirely new window into the weak interactions of the
top quark.  We eagerly await their discovery in Run II at the
Tevatron.

{\it Note added}: While completing this work, another paper on
single-top-quark production at hadron colliders appeared \cite{BBD}.
This paper advocates using a signal consisting of the decay products
of the top quark plus one or two additional jets, with two $b$ tags.
This is similar to our $s$-channel analysis, although we require one
and only one additional jet.

\section*{Acknowledgements}

\indent\indent We are grateful for conversations and correspondance
with D.~Amidei, R.~Frey, C.~Hill, T.~Liss, F.~Paige, S.~Parke,
M.~Seymour, T.~Sjostrand, and W.-M.~Yao. This work was supported in
part by Department of Energy grant DE-FG02-91ER40677.  Z.~S.~was
supported in part by a grant from the UIUC Campus Research Board.

\clearpage

\end{document}